\documentstyle[11pt,epsf]{article}
\newlength{\mytopmargin}
\newlength{\myleftmargin}
\renewcommand{\theequation}{\thesection.\arabic{equation}}
\begin{document}
\noindent
\begin{center}{  \Large\bf
Finite $N$ Fluctuation Formulas 
for Random Matrices}
\end{center}
\vspace{5mm}

\noindent
\begin{center} T.H.~Baker\footnote{email: tbaker@maths.mu.oz.au} and
P.J.~Forrester\footnote{email: matpjf@maths.mu.oz.au}\\[2mm]
{\it Department of Mathematics, University of Melbourne, \\
Parkville, Victoria 3052, Australia}
\end{center}
\vspace{.5cm}

\begin{quote}
For the Gaussian and Laguerre random matrix ensembles, the probability
density function (p.d.f.) for the linear statistic $\sum_{j=1}^N \Big (
x_j - \langle x \rangle \Big )$ is computed exactly and shown to
satisfy a central limit theorem as $N \to \infty$. For the circular
random matrix ensemble the p.d.f.'s for the linear statistics
${1 \over 2} \sum_{j=1}^N (\theta_j - \pi)$ and $- \sum_{j=1}^N
\log 2|\sin \theta_j/2|$ are calculated exactly by using a constant
term identity from the theory of the Selberg integral, and are also
shown to satisfy a central limit theorem as $N \to \infty$.
\end{quote}

\vspace{.5cm}

\section{INTRODUCTION}
A number of rigorous results have recently been established
regarding probability density functions and associated fluctuation
formulas for linear statistics in random matrix ensembles. We recall
that with $\lambda_1,\dots,\lambda_N$ denoting the eigenvalues of a
random matrix, a linear statistic is any stochastic function $A$
which can be written in the form $A=\sum_{j=1}^Na(\lambda_j)$.
Corresponding to the statistic $A$ is the probability function
(p.d.f)
\begin{equation}
P(u):=\prod_{j=1}^N\int_{-\infty}^\infty d\lambda_1\dots
\int_{-\infty}^\infty d\lambda_N \, \delta\left(u-\sum_{j=1}^Na(
\lambda_j)\right)W(\lambda_1,\dots,\lambda_N),
\end{equation}
where $W$ denotes the p.d.f. for the eigenvalue distribution, for
the event that $u=\sum_{j=1}^Na(\lambda_j)$. An evaluation of the
variance of $P(u)$ is referred to as a fluctuation formula, although
this term is sometimes used to refer to an evaluation of $P(u)$
itself.

When $a(x)=\chi_{[-l/2,l/2]}-\langle\chi_{[-l/2,l/2]}\rangle$,
where $\langle\ \rangle$ denotes the mean and $\chi_B=1$ if $x\in
B$, $\chi_B=0$ otherwise, we have that the linear statistic $A$
represents the deviation in the number of eigenvalues from the mean
number in the interval $[-l/2,l/2]$. For $N\times N$ Gaussian
random matrices, scaled so that the mean eigenvalue spacing in the
bulk of the spectrum is unity, it has been proved by Costin and
Lebowitz [1] that
\begin{equation}
P\left({\left({2\over{\pi^2\beta}}\log l\right)}^{1/2}u\right)\sim
e^{-u^2/2}\quad {\rm{as}}\:\: l\rightarrow\infty,
\end{equation}
where $\beta =1,2$ and 4 for random symmetric, Hermitian and
self-dual real quaternion matrices respectively. Before scaling, the
eigenvalue p.d.f. for Gaussian random matrices is proportional to
\begin{equation}
\prod_{l=1}^N e^{-\beta\lambda_l^2/2}\prod_{1\leq j<k\leq
N}{|\lambda_k-\lambda_j|}^\beta.
\end{equation}
This can be interpreted as the Boltzmann factor for a log-gas
system in equilibrium at inverse temperature $\beta$. From this
interpretation, heuristic arguments based on macroscopic
electrostatics have been devised [2,3] which support the validity
of (1.2) for all $\beta>0$.

Another, more general class of rigorous results for the evaluation
of (1.1) for random matrix ensembles has been obtained by Johansson
[4,5] (see also [6] in the case $\beta = 2$). In these results,
instead of scaling the eigenvalues so that the mean spacing is
unity, the scale is chosen so that as
$N\rightarrow\infty$ the support of the density of the eigenvalues
is the finite interval $(-1,1)$. Since for large $N$ the density of
eigenvalues $\rho(x)$ implied by (1.3) is given by the Wigner
semi-circle law (see e.g. ref. [7])
\begin{equation}
\rho(x)\sim{{{(2N)}^{1/2}}\over\pi}{\left(1-{{x^2}\over
{{(2N)}^{1/2}}}\right)}^{1/2},
\end{equation}
this is achieved by the scaling
$\lambda_j\mapsto{(2N)}^{1/2}\lambda_j$ in (1.3). More generally,
with the exponent $\lambda_l^2$ in (1.3) replaced by an even-degree
polynomial which is positive for large $\lambda_l$, a scale can
always be chosen so that the support is the finite interval $(-1,1)$
[4]. Moreover, independent of the details of the polynomial,
Johansson proved that the p.d.f.~(1.1) in the scaled
$N\rightarrow\infty$ limit tends to a Gaussian:
\setcounter{equation}{0}
\renewcommand{\theequation}{1.5\alph{equation}}
\begin{equation}
\lim_{N\rightarrow\infty\atop{\rm{support}}\
\rho(x)=(-1,1)} P(u)=e^{-u^2/2\sigma^2}
\end{equation}
where
\begin{equation}
\sigma^2={1\over{\beta\pi^2}}\int_{-1}^1dx{{a(x)}\over{{(1-x^2)}^{1/2}}}
\int_{-1}^1dy{{a'(y){(1-y^2)}^{1/2}}\over{x-y}}
\end{equation}
provided $\sigma^2$ is finite. The value of the variance was
predicted in earlier work due to Br\'ezin and Zee 
[8] in the case $\beta=2$, and Beenakker [9] for general $\beta>0$.
\setcounter{equation}{5}
\renewcommand{\theequation}{\thesection.\arabic{equation}}

Both (1.2) and (1.5) can be interpreted as central limit theorems.
However, unlike the classical result the standard deviation is no
longer proportional to $\sqrt{N}$ (or $\sqrt{l}$ in the case of
(1.2)). Rather the fluctuations are strongly suppressed, being
independent of $N$ altogether in the case of (1.5).

It is the purpose of this work to extend the rigorous results
relating to the p.d.f.~(1.1) for random matrix ensembles. Attention
will be focussed on two types of linear statistics: $a(x)=x$ and
$a(x)=-\log|x|$, which both arise naturally within the log-gas
interpretation of (1.3). Indeed $a(x)=x$ corresponds to the dipole
moment while $a(x)=-\log|x|$ corresponds to the potential at the
origin. This latter statistic has been considered for the
two-dimensional one-component plasma (i.e. the two-dimensional
generalization of the log-gas) by Alastuey and Jancovici [10].

In Section 2 we will show how (1.1) with $W$ proportional to (1.3)
can be computed exactly for $a(x)=x$. Also, we calculate (1.1)
exactly for $a(x)=x-\langle x\rangle$ with $W$ proportional to
\begin{equation}
\prod_{l=1}^N x_l^{\beta a/2}e^{-\beta x_l/2}\prod_{1\leq j<k\leq
N}{|x_k-x_j|}^\beta,
\end{equation}
$x_j \ge 0$ $(j=1,\dots,N)$,
corresponding to the Laguerre random matrix ensemble. In the scaled
$N\rightarrow\infty$ limit, the result (1.5) with $a(x)=x$ is
reclaimed in both cases.

In Section 3 we consider (1.1) with $W$ proportional to
\begin{equation}
\prod_{1\leq j<k\leq
N}{\left(2|\sin(\theta_k-\theta_j)/2|\right)}^\beta,
\end{equation}
$0 \le \theta_j \le 2 \pi$ $(j=1,\dots,N)$,
corresponding to Dyson's circular ensemble of unitary
random matrices. In the case $\beta=2$ the large-$N$ behaviour of
$P(u)$ for
$a(\theta)=(\theta-\pi)/2$ and $a(\theta)=-\log2|\sin\theta/2|$ is
computed using the so-called Fisher-Hartwig conjecture [11] (which
is now a theorem [12,13]) from the theory of Toeplitz determinants.
For general $\beta$, $P(u)$ is evaluated for these statistics by
using a constant term identity due to Morris [14] which is
equivalent to the well known Selberg integral [15]. We conclude in
Section 4 with an interpretation of the variance of the dipole
moment statistic calculated in Section 2 as a susceptibility in
macroscopic electrostatics.

\section{DIPOLE MOMENT STATISTIC IN THE GAUSSIAN AND LAGUERRE
ENSEMBLE}
\setcounter{equation}{0}
 Rather than study the p.d.f. (1.1) directly, we consider
instead its Fourier transform
\begin{eqnarray}
\tilde{P}(k)&:=&\int_{-\infty}^\infty dx \, e^{ikx}P(x) \nonumber \\
&=&\int_{-\infty}^\infty
dx_1 \, e^{ika(x_1)}\dots\int_{-\infty}^\infty
dx_N \, e^{ika(x_N)}W(x_1,\dots,x_N).
\end{eqnarray}
Note that (2.1) can be interpreted as the canonical average of the
Boltzmann factor for a one-body external potential $ika(x)/\beta$.

\subsection{Gaussian random matrices}

There are three distinct random matrix ensembles, in which the
matrices $X$ are real symmetric ($\beta=1$), Hermitian ($\beta=2$)
and self-dual quaternion real ($\beta=4$), and the joint distribution
for their elements is proportional to
$e^{-\beta{\rm{Tr}}(X^2)/2}$. The corresponding eigenvalue
p.d.f. is proportional to (1.3). Scaling the eigenvalues
$\lambda_j\mapsto\sqrt{2N}\lambda_j$ so that the support of the
density is $(-1,1)$, we have from (1.3) and (2.1) that for the
linear statistic corresponding to $a(x)=x$
\begin{eqnarray}
\tilde{P}(k)&=&\!{1\over C}\int_{-\infty}^\infty
d\lambda_1e^{-N\beta\lambda_1^2+ik\lambda_1}\dots
\int_{-\infty}^\infty
d\lambda_Ne^{-N\beta\lambda_N^2+ik\lambda_N}\prod_{1\leq j<k\leq
N}{|\lambda_k-\lambda_j|}^\beta \nonumber \\
&=&\! {{e^{-k^2/4\beta}}\over
C}\int_{-\infty}^\infty d\lambda_1e^{-N\beta{(\lambda_1-ik/2\beta
N)}^2}\dots\int_{-\infty}^\infty
d\lambda_Ne^{-N\beta{(\lambda_N-ik/2\beta N)}^2}\prod_{1\leq j<k\leq
N}{|\lambda_k-\lambda_j|}^\beta 
\end{eqnarray}
where the normalization $C$ is such that $\tilde{P}(0)=1$.

Now for $\beta$ even the integrand in (2.2) is analytic and decays
sufficiently fast at infinity that we can shift the contours of
integration from the real line to $\lambda_j=ik/2\beta N+t_j$
$(j=1,\dots,N)$, which gives that the integral is independent of
$k$. Since the integral in (2.2) divided by $C$ is a bounded
analytic function of $\beta$ for ${\rm{Re}}(\beta)>0$, it follows
from Carlson's theorem [16] that the integral is independent of $k$
for all ${\rm{Re}}(\beta)>0$. Thus for all $\beta>0$ and each
$N=1,2,\dots$ we have the simple result that
\setcounter{equation}{0}
\renewcommand{\theequation}{2.3\alph{equation}}
\begin{equation}
\tilde{P}(k)=e^{-k^2/4\beta}
\end{equation}
and so
\begin{equation}
P(u)={\left({1\over{\pi\beta}}\right)}^{1/2}e^{-\beta u^2}.
\end{equation}
Since (2.3b) is independent of $N$ it trivially remains valid in
the $N\rightarrow\infty$ limit, and agrees with (1.5) provided
$\sigma^2=1/2\beta$.

\setcounter{equation}{3}
\renewcommand{\theequation}{\thesection.\arabic{equation}}

The value of $\sigma^2$ is easily computed from (1.5b) by recalling
(see e.g. [17]) that the solution of the integral equation
\begin{equation}
x=\int_{-1}^1dy{{\phi(y)}\over{x-y}}
\end{equation}
is $\phi(y)={1\over\pi}\sqrt{1-y^2}$. Thus with $a(x)=x$,
substituting (2.4) in (1.5b) we have
\begin{equation}
\sigma^2={1\over{\beta\pi}}\int_{-1}^1{{x^2}\over{\sqrt{1-x^2}}}dx=
{1\over{2\beta}},
\end{equation}
as required.

In Fig.~1 we illustrate (2.3b) by empirically calculating $P(u)$
for the eigenvalues of 5,000 $2\times 2$ matrices from the Gaussian
Orthogonal Ensemble of random real symmetric matrices
($\beta=1$) and comparing the empirical p.d.f.~to the theoretical
value, eq.~(2.3b) with $\beta = 1$.
\subsection{Laguerre random matrix ensemble}
If $X$ is a random $n\times m$ ($n\geq m$) matrix with Gaussian
entries, which are either all real ($\beta=1$), complex ($\beta=2$)
or real quaternion ($\beta=4$) with joint distribution of the
elements proportional to $e^{-\beta{\rm{Tr}}(X^\dagger X)/2}$, then
the eigenvalue p.d.f. of $X^\dagger X$ is proportional to (1.3) with
$N=m$ and $a=n-m+\chi_\beta$ ($\chi_\beta=-1,0,1$ for
$\beta=1,2,4$ respectively). For this so called Laguerre ensemble
it is known (see e.g. [6]) that the support of the eigenvalue
density is $(0,4N)$. Scaling the eigenvalues
$\lambda_j\mapsto2N\lambda_j$, so that the support is (0,2) (the
important point here is that the length of the interval is 2, as is
required for the validity of (1.5b)), (2.1) with $a(x)=x-\langle
x\rangle$ reads
\begin{equation}
\tilde{P}(k)={1\over C}\int_0^\infty dx_1x_1^{\beta a/2}e^{-\beta
Nx_1+ik(x_1-\langle x\rangle)}\dots\int_0^\infty dx_Nx_N^{\beta a/2}
e^{-\beta Nx_N+ik(x_N-\langle x\rangle)}\hspace{-1\jot}\prod_{1\leq
j<k\leq N}{|x_k-x_j|}^\beta,
\end{equation}
where $C$ is such that $\tilde{P}(0)=1$ and
\begin{equation}
\langle x\rangle:={1\over C}\int_0^\infty dx_1 \, x_1^{\beta
a/2}e^{-\beta Nx_1}\dots\int_0^\infty dx_N \, x_N^{\beta
a/2}e^{-\beta Nx_N}\left({1 \over
N}\sum_{j=1}^Nx_j\right)\prod_{1\leq j<k\leq N}{|x_k-x_j|}^\beta.
\end{equation}

To evaluate (2.7), we note that a simple change of variables gives
\begin{eqnarray}
&&\int_0^\infty dx_1 \, x_1^\alpha e^{-\mu bx_1}\dots\int_0^\infty
dx_N \, x_N^\alpha e^{-\mu bx_N}\prod_{1\leq j<k\leq
N}{|x_k-x_j|}^\beta \nonumber \\
&&=\mu^{-N(\alpha+1)-\beta N(N-1)/2}\int_0^\infty
dx_1 \, x_1^\alpha e^{-bx_1}\dots\int_0^\infty dx_N \, x_N^\alpha
e^{-bx_N}\prod_{1\leq j<k\leq N}{|x_k-x_j|}^\beta,
\end{eqnarray}
and that the integral in (2.7) can be obtained from the left hand
side of (2.8) by differentiation with respect to $\mu$ and setting
$\alpha=\beta a/2$, $b=\beta N$ and $\mu=1$. Applying the same
operation to the right hand side of (2.8) gives
\begin{equation}
\langle x\rangle={1\over
2}+{1\over{2N}}\left(a-1+{2\over\beta}\right).
\end{equation}

Now, the non-trivial dependence on $k$ in (2.6) occurs only in the
factors $e^{-(\beta N+ik)x_j}$. This dependence can be taken
outside the integral by the change of variables $(\beta
N-ik)x_j=\beta Ny_j$ (this operation is immediately valid for
$\beta$ even; it remains valid for ${\rm{Re}}(\beta)>0$ by
Carlson's theorem) to give
\begin{equation}
\tilde{P}(k)=e^{-ikN\langle x\rangle}{\left({1\over{1-ik/\beta
N}}\right)}^{N^2\beta\langle x\rangle}.
\end{equation}
Taking the inverse transform we have
\begin{equation}
P(u)={1\over{\Gamma(\beta N^2\langle x\rangle)}}{(\beta N)}^{\beta
N^2\langle x\rangle}{(u+N\langle x\rangle)}^{\beta N^2\langle
x\rangle-1}e^{-\beta N(u+N\langle x\rangle)},
\end{equation}
for $u>-N\langle x\rangle$, $P(u)=0$ for $u<-N\langle x\rangle$. In
the limit $N\rightarrow\infty$ we see that (2.10) tends to (2.3a)
as expected, thus explicitly demonstrating the universality 
feature of (1.5).

In Fig.~2 we illustrate (2.11) by numerically calculating a
histogram for the p.d.f.~of $\sum_{j=1}^N ( \lambda_j - \langle
\lambda \rangle )$ for 3,000 matrices $X^T X$, where $X$ is a
$3 \times 2$ real rectanglar matrix with entries chosen with
p.d.f.~${1 \over \sqrt{2 \pi}} e^{-x_{jk}^2/2}$, and comparing it
against the theoretical prediction (2.11) with $\beta = 1$,
$N=2$ and $\langle x \rangle = 1/2 \, + \, 1/2N$.
\section{DIPOLE MOMENT AND POTENTIAL STATISTIC IN THE CIRCULAR
ENSEMBLE}
\setcounter{equation}{0}
\subsection{The variance}
For a general linear statistic $A$,
\begin{eqnarray}
\sigma^2&:=&\langle{(A-\langle A\rangle)}^2\rangle\nonumber \\
&=&\int_Idx\int_Idy \,a(x)a(y)\left(\rho_2^T(x,y)+\rho_1(x)
\delta(x-y)\right),
\end{eqnarray}
where $\rho_1(x)$ denotes the density, $\rho_2^T$ the truncated two
particle distribution and $I$ the allowed domain for the particles
(eigenvalues). For the circular ensemble we have $I=[0,2\pi)$ and due
to the periodicity we can write
\begin{equation}
\rho_2^T(x,y)+\rho_1(x)\delta(x-y)=\sum_{n=-\infty}^\infty
r_ne^{i(x-y)n}.
\end{equation}
Substituting (3.2) in (3.1) gives
\begin{equation}
\sigma^2=2\pi\sum_{n=-\infty}^\infty{|a_n|}^2r_n,\quad {\rm{where}}
\quad
a(x)=\sum_{n=-\infty}^\infty a_ne^{inx},\quad x\in[0,2\pi).
\end{equation}

Consider now the particular linear statistics with $a_1(x)={1\over
2}(x-\pi)$ and $a_2(x)=-\log2|\sin x/2|$. We have
\begin{equation}
a_1(x)=\sum_{m=1}^\infty{{\sin mx}\over m}\quad {\rm{and}}\quad
a_2(x)=\sum_{m=1}^\infty{{\cos mx}\over m},
\end{equation}
so to analyze $\sigma^2$ from the formula in (3.3) it remains to
specify the behavior of $r_n$. This can be done using an heuristic
electrostatic argument combined with linear response theory (see
e.g. [18]), which gives $r_n\sim|n|/\pi\beta$ for
$0\leq|n|\leq{\rm{O}}(N)$. Substituting in (3.4) we therefore have
\begin{equation}
\sigma^2\sim 2\pi\sum_{n=-cN \atop n \ne 0}^{cN}{1\over{\pi\beta
2^2|n|}}\sim{1\over\beta}\log N
\end{equation}
for both the dipole moment and potential statistics. Linear response
arguments [19,20] also give the prediction that in an appropriate
macroscopic limit (which here corresponds to $N\rightarrow\infty$)
the p.d.f. of any linear statistic will be Gaussian, thus
suggesting that for the dipole moment and potential statistics 
\begin{equation}
P\left({\left({1\over\beta}\log
N\right)}^{1/2}u\right)\sim{\left({1\over{2\pi}}\right)}^{1/2}
e^{-u^2/2}\quad  {\rm{as}}\quad  N\rightarrow\infty.
\end{equation}
\subsection{The case $\beta=2$}
>From the Vandermonde determinant expansion
\begin{equation}
{\rm{det}}\left[ z_j^{k-1}\right]_{j,k=1,\dots,N}=\prod_{1\leq
j<k\leq N}(z_k-z_j)
\end{equation}
it is straightforward to show (see e.g. [21]) that with $W$ given by
(1.7) and $\beta=2$ (2.1) can be rewritten in terms of a Toeplitz
determinant:
\begin{equation}
\tilde{P}(k)=D_N\left[
e^{ika(\theta)}\right]:={\rm{det}}\left[{1\over{2\pi}}\int_0^{2\pi}
\, d\theta
e^{ika(\theta)}e^{i\theta(j-k)}\right]_{j,k=1,\dots,N}.
\end{equation}
Using an asymptotic formula for the large-$N$ form of $D_N$, known
as the Fisher-Hartwig conjecture [11-13], we can rigorously establish
(3.6).

\vspace{.2cm}
\noindent
{\bf Proposition 3.1} (Fisher-Hartwig conjecture)
In (3.8) let
\begin{equation}
ika(\theta)=g(\theta)-i\sum_{r=1}^Rb_r(\pi-(\theta-\theta_r))+
\sum_{r=1}^Ra_r\log|2-2\cos(\theta-\theta_r)|
\end{equation}
and assume $g(\theta)=\sum_{p=-\infty}^\infty g_pe^{i\theta}$ where
$\sum_{p=-\infty}^\infty|p|{|g_p|}^2<\infty$. Then for
${\rm{Re}}(\alpha_r)>-1/2$ and ${\rm{Re}}(\beta_r)=0$
\begin{equation}
D_N\left[e^{ika(\theta)}\right]\sim e^{g_0
N}e^{\sum_{r=1}^R(a_r^2-b_r^2)\log N}E
\end{equation}
where $E$ is independent of $N$. To specify $E$, write
$g(\theta)-g_0=g_+(e^{i\theta})+g_-(e^{-i\theta})$ where
$g_+(z)=\sum_{p=1}^\infty g_pz^p$ and
$g_-(z)=\sum_{p=-\infty}^{-1}g_pz^p$. Then
\begin{eqnarray}
&&E=e^{\sum_{k=1}^\infty
jg_kg_{-k}}\prod_{r=1}^Re^{-(a_r+b_r)g_-(e^{i\theta_r})}
e^{-(a_r-b_r)g_+(e^{-i\theta})}\nonumber \\
&&\times\prod_{1\leq r<s\leq
R}{|1-e^{i(\theta_s-\theta_r)}|}^{-(a_r+b_r)(a_s-b_s)}
\prod_{r=1}^R{{G(1+a_r+b_r)G(1+a_r-b_r)}\over{G(1+2a_r)}},
\end{eqnarray}
where $G$ is the Barnes $G$-function defined by
\begin{equation}
G(z+1)={(2\pi)}^{z/2}\exp\left(-z/2-(\gamma+1)z^2/2\right)
\prod_{k=1}^\infty{\left(1+{z\over k}\right)}^k\exp(-z+z^2/k)
\end{equation}
($\gamma$ denotes Euler's constant), which
 has the special values
$G(1)=G(2)=1$ and satisfies the functional relation
\begin{equation}
G(z+1)=\Gamma(z)G(z).
\end{equation}

\vspace{.2cm}

First consider the application of (3.9) to the calculation of the
p.d.f.~for the potential statistic. From (3.8) we have
\begin{equation}
\tilde{P}(k)=D_N\left[{(2-2\cos\theta)}^{ik/2}\right],
\end{equation}
so we take $g(\theta)=0$, $R=1$, $b_1=0$, $a_1={{ik}\over
2}{({1\over 2}\log N)}^{-1/2}$ and $\theta_1=0$ in (3.9) and (3.10)
to conclude that
\begin{equation}
\tilde{P}\left({k\over{{({1\over 2}\log N)}^{1/2}}}\right)\sim
e^{-k^2/2}
\end{equation}
which is equivalent to (3.6) with $\beta=2$.

For the dipole moment statistic we have from (3.8) that
\begin{equation}
\tilde{P}(k)=D_N\left[e^{-{{ik}\over2}(\pi-\theta)}\right].
\end{equation}
Thus we take $g(\theta)=0$, $R=1$, $a_1=0$, $b_1={k\over 2}$ and
$\theta_1=0$ in (3.9) and (3.10) to conclude that (3.15) again
holds, as predicted from (3.6).

\subsection{General $\beta$}
Here we will show that $\tilde{P}(k)$ for the dipole moment and
potential statistics in the circular ensemble can be given in
closed form for general $\beta$, and prove that the expected
asymptotic behaviur (3.6) is valid for general rational $\beta$ (at
least). Our chief tool for this purpose is a constant term identity
of Morris [14], written as the multidimensional integral evaluation
[22]
\begin{eqnarray}
M_n(a,b,c)&:=&\prod_{l=1}^n{1\over{2\pi}}\int_{-\pi}^\pi d\theta_l
e^{i(a-b)\theta_l/2}{|1+e^{i\theta_l}|}^{a+b}\prod_{1\leq j<k\leq
n}{|e^{i\theta_k}-e^{i\theta_j}|}^{2c}\nonumber \\
&=&\prod_{j=0}^{n-1}{{\Gamma(c(j+1)+1)\Gamma(cj+a+b+1)}
\over{\Gamma(cj+a+1)\Gamma(cj+b+1)\Gamma(1+c)}}.
\end{eqnarray}

First consider the dipole moment statistic. From (1.7) and (2.1) we
have
\begin{eqnarray}
\tilde{P}(k)&=&{1\over
C}\prod_{l=1}^N\int_0^{2\pi}d\theta_le^{ik(\theta_l-\pi)/2}\prod_{1\leq
j<k\leq N}{|e^{i\theta_k}-e^{i\theta_j}|}^\beta \nonumber \\
&=&{{M_N(k/2,-k/2,\beta/2)}\over{M_N(0,0,\beta/2)}}=\prod_{j=0}^{N-1}
{{{(\Gamma(\beta j/2+1))}^2}\over{\Gamma(\beta j/2+k/2+1)
\Gamma(\beta j/2-k/2+1)}}.
\end{eqnarray}
Although (3.18) is a closed form evaluation, it is not convenient
for the determination of the large-$N$ asymptotics. This same
problem has been faced in an earlier application of the Morris
constant term identity [23]. Its resolution is to make use of the
identity
\begin{equation}
\prod_{j=0}^{N-1}\Gamma(\alpha+1+jc)=c^{cN(N-1)/2+N(\alpha+1/2)}
{(2\pi)}^{-N(c-1)/2}\prod_{l=1}^c{{G(N+(\alpha+l)/c)}
\over{G((\alpha+l)/c)}},
\end{equation}
valid for $c$ a positive integer. We can use (3.19) in (3.18) for all
rational
$\beta$, $\beta/2=s/r$ with $s$ and $r$ relatively prime positive
integers say. This is done by replacing $N$ by $rN$ and noting
\begin{equation}
\prod_{j=0}^{rN-1}\Gamma\left(\alpha+1+{{js}\over
r}\right)=\prod_{p=0}^{r-1}\prod_{k=0}^{N-1}\Gamma\left(
\alpha+1+{{ps}\over r}+sk\right).
\end{equation}
Thus for $\beta/2=s/r$ and with $N$ replaced by $rN$, (3.18) can be
rewritten to read
\begin{equation}
\tilde{P}(k)=\prod_{p=0}^{r-1}\prod_{l=1}^s{{{(G(N+ps/r+l/s))}^2
G(ps/r+(l+k/2)/s)G(ps/r+(l-k/2)/s)}
\over{G(N+ps/r+(l+k/2)/s)G(N+ps/r+(l-k/2)/s){(G(l/s+ps/r))}^2}}.
\end{equation}

The large-$N$ asymptotics of $\tilde{P}(k)$ can be deduced from
(3.21) by making use of the asymptotic formula [24]
\begin{equation}
\log\left({{G(N+a+1)}\over{G(N+b+1)}}\right)\sim(b-a)N+{{a-b}\over
2}\log 2\pi+\left((a-b)N+{{a^2-b^2}\over 2}\right)\log
N+{\rm{o}}(1).
\end{equation}
This gives
\begin{equation}
\log\tilde{P}(k)\sim-{{k^2}\over{2\beta}}\log
N+\sum_{p=0}^{r-1}\sum_{l=1}^s\log{{G(ps/r+(l+k/2)/s)G(ps/r+(l-k/2)/s)}
\over{{(G(l/s+ps/r))}^2}}+{\rm{o}}(1),
\end{equation}
which implies
\begin{equation}
\tilde{P}\left({k\over{{\left({1\over\beta}\log
N\right)}^{1/2}}}\right)\sim e^{-k^2/2}
\end{equation}
in precise agreement with the Fourier transform of the expected
asymptotic behaviour (3.6).

Let us now turn our attention to the potential statistic. In this
case, from (2.1), (1.7) and (3.17) we have
\begin{eqnarray}
\tilde{P}(k)&=&{1\over
C}\prod_{l=1}^N\int_0^{2\pi}d\theta_l \, {|1-e^{i\theta_l}|}^{-ik}
\prod_{1\leq j<k\leq N}{|e^{i\theta_k}-e^{i\theta_j}|}^\beta
\nonumber \\ &=&{{M_N(-ik/2,-ik/2,\beta/2)}\over{M_N(0,0,\beta/2)}}=
\prod_{j=0}^{N-1}{{\Gamma(\beta
j/2-ik+1)\Gamma(\beta j/2+1)}\over{{(\Gamma(\beta j/2-ik/2+1))}^2}}
\end{eqnarray}
Proceeding as above, for rational $\beta$ ($\beta/2=s/r$), by using
(3.19) and (3.20) we can rewrite (3.25) as
\begin{equation}
\tilde{P}(k)=\prod_{p=0}^{r-1}\prod_{l=1}^s{{G(N+ps/r+(l-ik)/s)
G(N+ps/r+l/s){(G(ps/r+(l-ik/2)/s))}^2}\over
{{(G(N+ps/r+(l-ik/2)/s))}^2G(ps/r+(l-ik)/s)G(ps/r+l/s)}}.
\end{equation}
The asymptotic formula (3.22) now gives that for
$N\rightarrow\infty$
\begin{equation}
\log\tilde{P}(k)\sim-{{k^2}\over{2\beta}}\log
N+\sum_{p=0}^{r-1}\sum_{l=1}^s\log{{{(G(ps/r+(l-ik/2)/s))}^2}\over
{G(ps/r+(l-ik)/s)G(ps/r+l/s)}}+{\rm{o}}(1)
\end{equation}
which again implies the anticipated result
(3.24).

In Fig.~3 we illustrate (3.18) by numerically constructing a
histogram for the p.d.f.~of ${1 \over 2} \sum_{j=1}^N ( \theta_j -
\pi)$ for 5,000 $3 \times 3$ matrices from the CUE (these are
constructed according to the procedure specified in [25]),
and comparing it against the inverse transform of (3.18) in the
case $N=3$, $\beta = 2$.
\section{RELATIONSHIP BETWEEN THE VARIANCE OF THE DIPOLE MOMENT
STATISTIC AND THE SUSCEPTIBILITY}
\setcounter{equation}{0}
In this final section we will show how the result (2.5) can be
anticipated from its interpretation as a susceptibility. We recall
that in macroscopic electrostatics the susceptibility tensor $\chi$
relates the electric polarization density of a Coulomb system,
confined to a region $\Lambda$ in a vacuum, to the applied electric
field. The laws of macroscopic electrostatics allow the components
of $\chi$ to be expressed in terms of the dielectric constant of
the system. In particular, for a conducting ellipse, this theory
gives [26]
\begin{equation}
\chi_{xx}={1\over{2\pi}}{{L_x+L_y}\over{L_x}}
\end{equation}
where $L_x$ is the length and $L_y$ the width of the ellipse.

On the other hand, the susceptibility can be related to the
microscopic quantities by linear response theory. With the electric
field applied in the $x$-direction, this approach gives [27]
\begin{equation}
\chi_{xx}={\beta\over{|\Lambda|}}\left(\langle
P_x^2\rangle-{\langle P_x\rangle}^2\right)
\end{equation}
where $P_x$ is the $x$-component of the instantaneous polarization
(or equivalently, dipole moment) $\vec{P}:=\sum_jq_j\vec{r_j}$.

Comparing (4.1) and (4.2), and noting that $|\Lambda|={\pi\over
4}L_xL_y$ we obtain
\begin{equation}
\langle P_x^2\rangle-{\langle
P_x\rangle}^2={1\over{8\beta}}L_x(L_x+L_y).
\end{equation}
Now we construct an interval of length 2 in the $x$-direction from
the ellipse by setting $L_x=2$, $L_y=0$. This gives
\begin{equation}
\langle P_x^2\rangle-{\langle P_x\rangle}^2={1\over{2\beta}}
\end{equation}
which is in precise agreement with the result (2.5) for the
variance of the dipole moment statistic.

\section*{ACKNOWLEDGEMENTS}
This work was supported by the Australian Research Council.

\pagebreak
\section*{FIGURE CAPTIONS}

\noindent
{\bf Figure 1.} \quad Empirical and theoretical p.d.f.~for the
statistic $\sum_{j=1}^N \lambda_j$, where the $\lambda_j$ are
scaled eigenvalues of $2 \times 2$ GOE matrices.

\vspace{.2cm}
\noindent
{\bf Figure 2.} \quad Empirical and theoretical p.d.f.~for the
statistic $\sum_{j=1}^N \Big ( x_j - \langle x \rangle \Big )$
where the $x_j$ are scaled eigenvalues for the matrix $X^T X$.
Here $X$ is a $3 \times 2$ real rectangular random matrix with
Gaussian entries.

\vspace{.2cm}
\noindent
{\bf Figure 3.} \quad Empirical and theoretical p.d.f.~for the
statistic ${1 \over 2}\sum_{j=1}^N \Big (\theta_j - \pi \Big )$,
where the $\theta_j$ $(0 \le \theta_j < 2 \pi)$ are the phases for
the eigenvalues of $3 \times 3$ CUE matrices.

\pagebreak
\section*{REFERENCES}
\begin{description}
\item[][1] O.~Costin and J.L. Lebowitz, Phys. Rev. Lett. {\bf 75}, 69
(1995)
\item[][2] F.J.~Dyson, {\it The Coulomb fluid and the fifth Painlev\'e
transcendent}, preprint
\item[][3] M.M.~Fogler and B.I.~Shklovskii, Phys.~Rev.~Lett. {\bf 74},
3312 (1995)
\item[][4] K.~Johansson, 
{\it On fluctuations of random Hermitian matrices},
preprint
\item[][5] K.~Johansson, {\it On random matrices associated with compact
classical groups}, preprint
\item[][6] E.L.~Basor, preprint
\item[][7] T.H. Baker and P.J. Forrester, solv-int/9608004
\item[][8] E.~Br\'ezin and Z.A.~Zee, Nucl.~Phys. B {\bf 402}, 613 (1993)
\item[][9] C.W.J.~Beenakker, Nucl.~Phys.~B {\bf 422}, 515 (1994)
\item[][10] A.~Alastuey and B.~Jancovici, J.~Stat.~Phys. {\bf 34},
557 (1984)
\item[][11] M.E.~Fisher and R.E.~Hartwig, Adv.~Chem.~Phys. {\bf 15},
333 (1968)
\item[][12] A.~B\"ottcher and B.~Silbermann, {\it Analysis of Toeplitz
operators} (Berlin:Akademie, 1989)
\item[][13] E.L.~Basor and C.A.~Tracy, Physica A {\bf 177}, 167 (1991)
\item[][14] W.G.~Morris, {\it Constant term identities for finite and affine
root systems}, Ph.~D. thesis, University of Wisconsin, Madison (1982)
\item[][15] A.~Selberg, Norsk Mat.~Tids. {\bf 26}, 167 (1944)
\item[][16] E.C.~Titchmarsh, {\it Theory of Functions}, p.~186 (Oxford
University Press, London, 1939)
\item[][17] E.~Br\'ezin, C.~Itzykson, G.~Parisi and J.B.~Zuber, Commun.
Math. Phys., {\bf 59}, 35 (1978)
\item[][18] P.J.~Forrester, Nucl.~Phys.~B {\bf 435}, 421 (1995)
\item[][19] H.D.~Politzer, Phys.~Rev.~B {\bf 40}, 11917 (1989)
\item[][20] B.~Jancovici, J.~Stat.~Phys. {\bf 80}, 445 (1995)
\item[][21] G.~Szeg\"o, {\it Orthogonal Polynomials},
(AMS, Providence RI, 1967)
\item[][22] P.J.~Forrester, Mod.~Phys.~Lett B {\bf 9}, 359 (1995)
\item[][23] P.J.~Forrester, Physics Lett. A {\bf 163}, 121 (1992)
\item[][24] E.W.~Barnes, Quarterly J.~Pure Appl.~Math. {\bf 31}, 264
(1900)
\item[][25] K.~Zyczkowski and M.~Kus, J.~Phys.~A {\bf 27},
4235 (1994) 
\item[][26] Ph.~Choquard, B.~Piller and R.~Rentsch, J.~Stat.~Phys.
{\bf 46}, 599 (1987)
 \item[][27] Ph.~Choquard, B.~Piller and R.~Rentsch, J.~Stat.~Phys.
{\bf 43}, 197 (1986)
\end{description}

\begin{figure}
\epsfxsize=16cm
\epsfysize=12cm
\epsfbox{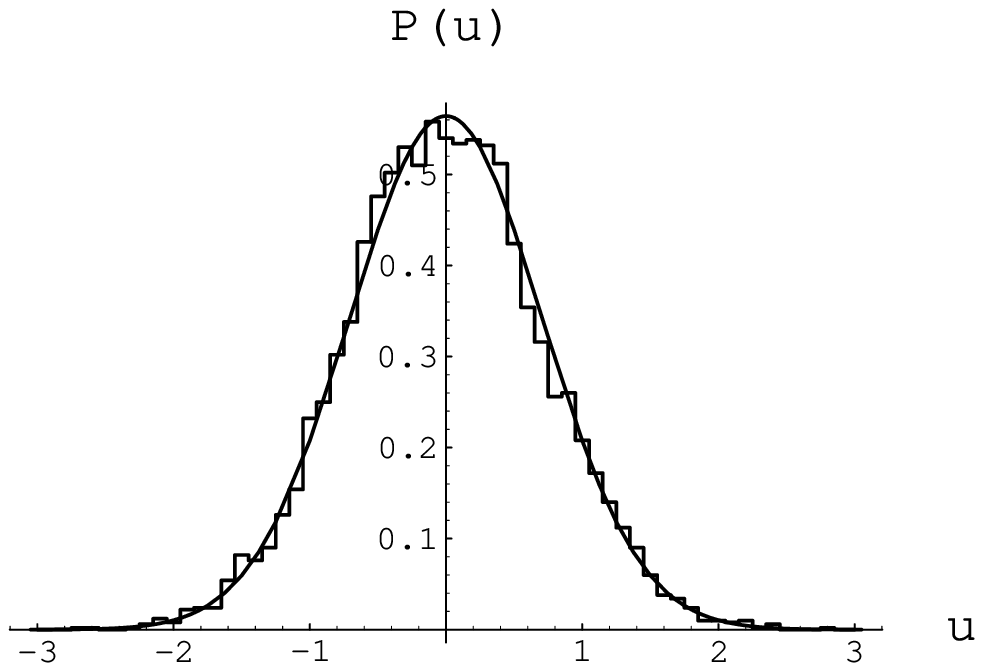}
\caption{ }
\end{figure}
\begin{figure}
\epsfxsize=16cm
\epsfysize=12cm
\epsfbox{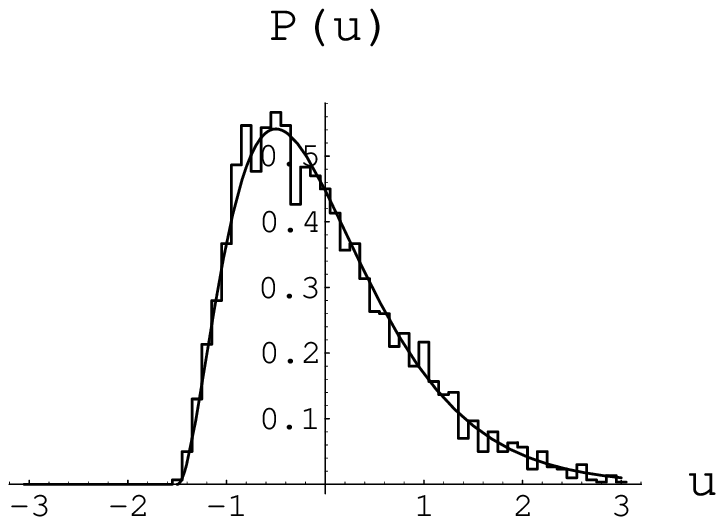}
\vspace*{1cm}
\caption{ }
\end{figure}
\begin{figure}
\epsfxsize=16cm
\epsfysize=12cm
\epsfbox{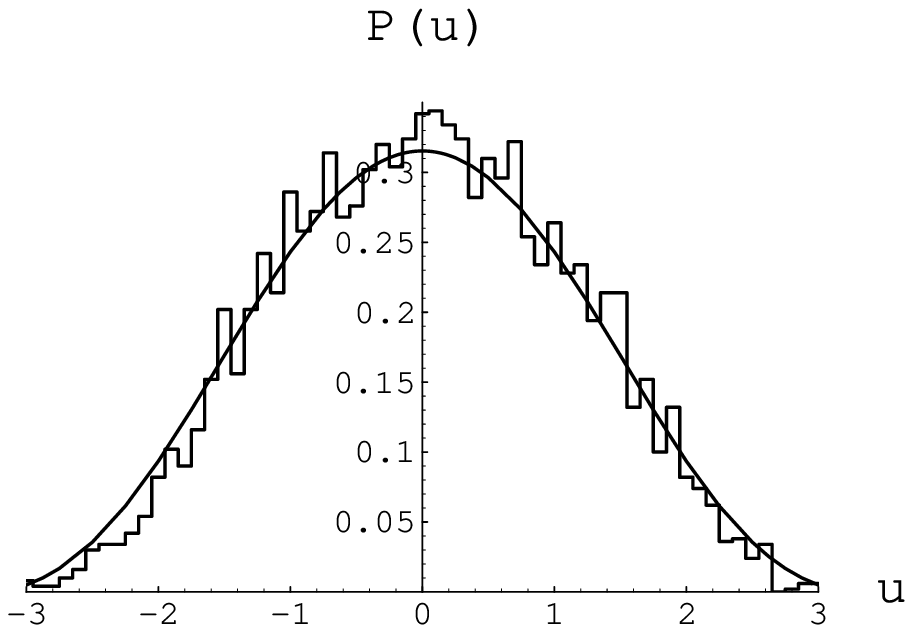}
\caption{ }
\end{figure}

\end{document}